\documentclass[prl,amsmath,amssymb,showpacs,showkeys,twocolumn]{revtex4}
\usepackage{graphicx}
\usepackage{dcolumn}
\usepackage{bm}
\usepackage{ifpdf}

      \ifpdf
      \usepackage[pdftex]{graphicx}
      \else
      \usepackage{graphicx}
     \fi

\begin{document}

\title{NMR of superfluid $^3$He in anisotropic aerogel}

\author{T.~Kunimatsu$^2$}
\author{T.~Sato$^1$}
\author{K.~Izumina$^1$}
\author{A.~Matsubara$^{124}$}
\author{Y.~Sasaki$^{24}$}
\author{M.~Kubota$^1$}
\author{O.~Ishikawa$^{3}$}
\author{T.~Mizusaki$^{24}$}
\author{Yu.~M.~Bunkov$^{15}$\/\thanks{e-mail: yuriy.bunkov@grenoble.cnrs.fr},}

\affiliation{
$^1$Institute for Solid State Physics, The University of Tokyo, Chiba 277-8581, Japan \\
$^2$Department of Physics, Graduate School of Science, Kyoto University, Kyoto 606-8502, Japan \\
$^3$Graduate School of Science, Osaka City University, Osaka 558-8585, Japan \\
$^4$Research Center for Low Temperature and Materials Sciences, Kyoto University, Kyoto 606-8502, Japan \\
$^5$Centre de Recherches sur les Tr\`es Basses Temp\'eratures, CNRS,
 38042, Grenoble, France}

\author{}
\address{}
\date{\today}
\begin{abstract}
We report on orientation of the order parameter in the $^3$He-A and
$^3$He-B phases caused  by aerogel anisotropy. In $^3$He-A we have
observed relatively homogeneous NMR line with an anomalously large
negative frequency shift. We can attribute this effect to an
orientation of orbital momentum along the axis of density
anisotropy. The similar orientation effect we have seen in $^3$He-B.
We can measure the A-phase Leggett frequency, which shows the same
energy gap suppression as in the B-phase. We observe a correlation
of A - B  transition temperature and NMR frequency shift.

\end{abstract}



\maketitle

\section{Introduction}
Pure superfluid $^3$He is certainly one of the most complex systems
in condensed matter which can be successfully described by a
comprehensive theory. The influence of disorder on ordered states is
one of the most interesting and ubiquitous problems in condensed
matter physics. In the case of superfluid $^3$He the disorder can be
produced by its impregnation in high porosity silica aerogel. It was
found, that the aerogel disorder leads to significant changes of
phase diagram of superfluid $^3$He. Not only the critical
temperature changes, but also the ground states of A and B phases as
well the temperature and dynamics of transition between them.  There
is not yet conclusive description of A like state in aerogel. It can
be "Robust" or "No-Robust" state  \cite{FominR,VolovikR}.
D.D.Osheroff \cite{Osheroff} suggested that the A-like phase is a
"Planar" phase, while Kyoto theoretical group \cite{KyoT} suggested
a "Polar" like state for anisotropic aerogel.

Particularly interesting is the question of the influence of a local
random anisotropy of aerogel on the  superfluid $^3$He-A order
parameter. The orbital part of the order parameter suppose to be
very sensitive to the anisotropy of aerogel density. The short scale
anisotropy can leads to a Larkin-Imry-Ma state as shown in a resent
work by Volovik \cite{VolovikA}. The anisotropy of scale longer then
texture healing length can leads to a texture and NMR inhomogeneity.
The global anisotropy can orients the order parameter throw out of
the sample. The question of natural aerogel anisotropy is now under
intensive investigations \cite{Halp}. The global anisotropy can be
achieved by aerogel deformation owing to the uniaxial pressure and
can lead to appearing a new phase, as suggested in \cite{KyoT}.

A related problem is the formation of topological defects and
disorder after a broken symmetry transition
\cite{volovik,grenoble-nature}. It was thought that in bulk $^3$He
the disorder and topological defects can easily disappears after
transition. In fact, some types of defects rest inside the
superfluid $^3$He even in the bulk conditions
\cite{grenoble-topdif}. In the superfluid $^3$He in aerogel the
topological defects can be pinned. This provides an interesting
example of a system with continuous symmetry in the presence of
random anisotropy disorder. It was suggested to describe it, at
least $^3$He-A, as the the Larkin-Imry-Ma state, contaminated  by
the network of the topological defects pinned by aerogel
\cite{VolovikA}. The experimental evidence of topological defects
contamination in $^3$He-B have been found  in Grenoble
\cite{Grendif}. There was shown that the NMR signal is a
superposition of the signals with different properties. Similar
observation for $^3$He-A was recently described in ref.
\cite{Mosdif}.

Both these problems are illuminated by investigations we presented
in this article. First of all we pay attention to the fact that the
orientation of orbital part of order parameter is very sensitive to
the density anisotropy of aerogel. Its oriented along a direction,
in which the concentration of aerogel is bigger. According to
 G. Volovik estimations \cite{VolovikA,VolovikE} the aerogel linear
 deformation of order of 0.1\% can  leads to a regular orientation
 of the order of parameter in $^3$He-A. We have
performed our experiments with superfluid $^3$He in aerogel under an
axial deformation, and found that the orbital momentum of superfluid
A and B phases is oriented along the axis of deformation. In result
of this orientation, the NMR of A phase display a large negative
frequency shift, while the NMR signal from B phase remains in
vicinity of the Larmor frequency. We can see the broadening of NMR
line, which is likely the effect of inhomogeneity of aerogel local
density orientation. Owing the negative frequency shift for A phase,
we was able, in a first time, to observe the dependance of A-B
transition temperature on the NMR frequency shift. This effect can
bring us a new ideas about mechanism of A-B transition in aerogel.
We have seen a very small influence of rotation on the NMR signals,
and only in the case, when we cross $T_c$ under fast rotation. We
can suggest that the main part of aerogel is contaminated by the
network of the topological defects pinned by aerogel.

\section{Experimental setup}

Our experiments were done at pressures 29.3 bar  in magnetic field
of 290 gauss, corresponding to NMR frequency of 940 kHz.  We used
the aerogel sample  of 98\% porosity aerogel in a form of cylinder
(diameter=5mm, length=3 mm) with the axis oriented along the
external steady magnetic field. The sample was kindly made by N.
Mulders. It was placed inside the Stycast 1266 epoxy cell. In the
first set of the experiments the aerogel fit loosely in the cell
except for a some force applied by the top and bottom of the cell,
which deformed the aerogel in the direction of the magnetic field
and removed any gap between the aerogel and top and bottom of the
cell. With cooling the Stycast body constrain on about 1\% more then
aerogel itself. Consequently, we can suppose that the aerogel sample
is deformed in a few percent in a direction of external magnetic
field. In a second set of experiments the additional Staycast
constrain was removed. By analyze of our results, we can conclude,
that the global orientation remains, but decrease. The influence of
Stycast constrain on order parameter orientation is clearly seen.

The cell was connected to the rest of $^3$He by a channel of 1 mm.
in diameter. The thermometry was done on the basis of a melting
curve thermometer. In order to avoid the formation of solid $^3$He
on the surface of aerogel strands, we have preplated aerogel by
$^4$He. The cell was installed on a rotating nuclear demagnetization
cryostat of ISSP, Tokyo university. Its give us ability to check the
influence of rotation on the observed NMR signals. We have used the
usual CW NMR spectroscopy as well the Homogeneous Precessing Domain
(HPD) method of NMR \cite{Bunrev}.

\section{Experimental results}

The NMR signals, we have observed at cooling, is shown in Fig.1. In
comparison with previous publications, we have observed in A-like
phase a relatively narrow NMR line with a big negative NMR frequency
shift. We can attribute it to the global orientation of orbital
momentum in $^3$He-A along the magnetic field. Below 0.82 of the
transition temperature in aerogel ($T_c^a$) the part of $^3$He make
a transition to B phase. We have found that at the conditions of our
experiments the positive NMR shift of $^3$He-B is very small. This
results show that orbital momentum in $^3$He-B also globally
oriented along the magnetic field! In Fig.2 we show the frequency
shift for the maximum of the NMR lines as well the left edges of the
line for A like and B like phases. We are also shows the frequency
shift for the second run, when the cell pressure was released.
\begin{figure}[ht]
\includegraphics[width=0.35\textwidth]{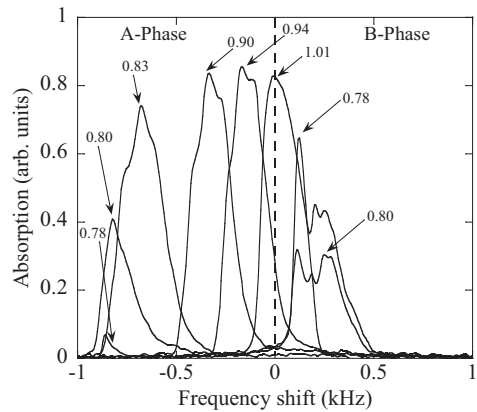}
\vspace{0.5cm}
 \caption{The signals of CW NMR for different temperatures
 (in units $T_c^a$) shown by number for each curve.
 The signals with negative frequency shift corresponds to $^3$He-A
 and with positive frequency shift - to $^3$He-B.}
 \label{cell}
\end{figure}
\begin{figure}[htb]
 \includegraphics[width=0.35\textwidth]{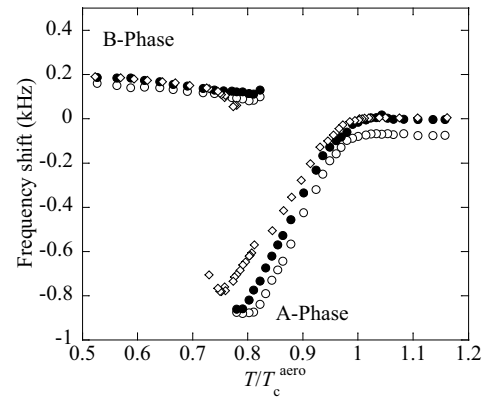}
 \caption{Position of the left edge ($\circ$) and the maximum
  ($\bullet$) of NMR line for the run with pressure and maximum
  ($\diamond$) for the run without additional pressure.}
 \label{collapse}
\end{figure}

The  transverse NMR frequency in A phase in the bulk depends from
the orientation of orbital momentum vector $\bf L$ and spin vector
$\bf d$.
\begin{equation}
\omega^2 = \omega_l^2 + \Omega_A^2 cos(2\phi) ,
\end{equation}
where $\omega_l$ is the Larmor frequency, $\Omega_A$ is the Leggett
 frequency in A phase and $\phi$ is the angle between $\bf L$ and
$\bf d$. At the magnetic fields we have applied, the spin vector
$\bf d$ should be oriented perpendicular to $\bf H$. The maximum
negative NMR frequency shift corresponds to the orientation of  $\bf
L$ along the magnetic field. If we have reached the vertical
orientation of $\bf L$, and if A like phase in aerogel is ABM state,
then the ratio of $\Omega_B$/$\Omega_A$ for the aerogel should be
the same as  for the bulk case.
 The $\Omega_B$  in aerogel have been studied
by different methods in \cite{DmitOB}. From these data we can
estimate $\Omega_B$ in aerogel at 29.3 as a 0.5 of $\Omega_B$ in a
bulk.  In Fig.3 we show the $\Omega_A$ in a bulk by dashed line, and
its 0.5 value by dotted line. We have found that at low temperatures
the measured $\Omega_A$ ($\bullet $) are approaches to dotted line
and corresponds to ABM state in aerogel. At a temperatures above 0.9
$T_c^a$ the measured $\Omega_A$ is significantly low. Can it be the
error of the method?  Part of NMR shift can be hidden by NMR
broadening. To estimate this, we have calculated $\Omega_A$ as a
shift from the center of a normal $^3$He NMR line to a left edge of
A-phase line ($\circ$). It is clearly seen that this effect can
explain the deviation only near $T_c^a$. At the temperature region
from 0.97 to 0.88 $T_c^a$ the deviation of $\Omega_A$ from the value
suggested by ABM state is clearly seen!  It may be that the order
parameter changes smoothly from planar to ABM state at cooling. This
our observation would be interesting to check by other methods.

\begin{figure}[htb]
\includegraphics[width=0.35\textwidth]{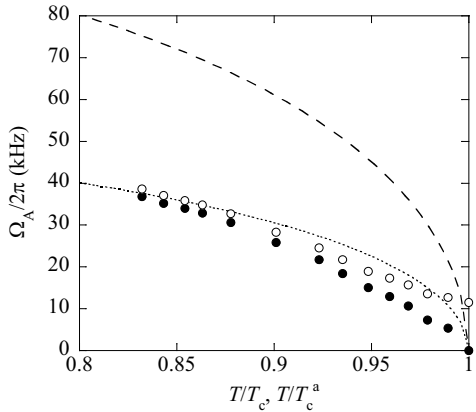}
 \caption{ The $\Omega_A$ calculated by the NMR shift from the left edge
 ($\bullet$) and from the maximum of NMR line ($\circ$) of normal $^3$He.
 The dashed line shows the value of $\Omega_A$ for bulk $^3$He,
 shown in  temperature scale of $T_c$ bulk. Dotted line, the half of $\Omega_A$
  for bulk shown in scale of $T_c$ in aerogel.}
 \label{f(b)}
\end{figure}

We are clearly seen that the transition temperature from A to B
phase depends on the frequency shift in $^3$He-A!  Previous studies
show that the A and B phases can coexist in some range of
temperatures \cite{DmitCO,HalpCO,OsCO}. But there was not known,
what parameter controls the temperature of this transition for
different parts of the sample. We found that at cooling the part of
$^3$He-A with larger frequency shift going to transition in a first.
(See Fig.1 and Fig.2) For the second set of experiments without
Stycast constrain the effect is even more profound. The bigger angle
of deviation of $\bf L$ from the magnetic field, the smaller
frequency shift and lower the temperature of transition. We have
succeed to observe this effect owing the negative frequency shift of
NMR. In a usual case of positive NMR shift for A phase it was
difficult to separate well A and B phase signals. We call
theoreticians for to explain the dependance of A-B transition
temperature from $\bf L$ orientation. The alternative explanation
can be made, if one suggest the distribution of aerogel density. The
regions with higher density can have lower $T_c^a$ lower temperature
of A-B transition and smaller negative frequency shift. For to
explain our results from a single run it should be of the order of
10\%, that is possible. But with the anisotropic deformation on 1\%
its should be also changed on about 10\%, (see Fig.2) which
completely exclude this explanation.
\begin{figure}[htb]
\includegraphics[width=0.35\textwidth]{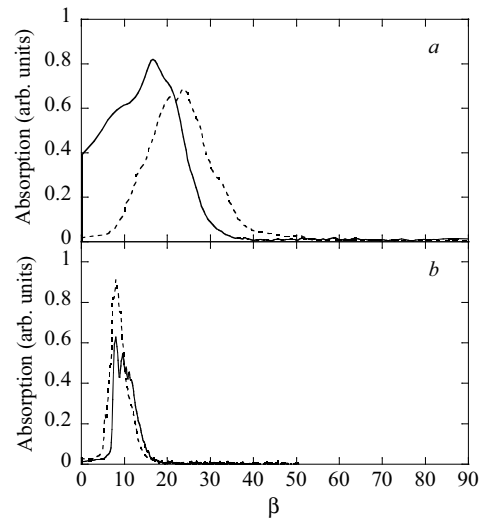}
 \caption{The distribution of NMR signal in the A (a) and in the B (b)
 phases presented as function of angle of deflection $\bf L$
 from a vertical direction in a local NMR approximation.
 The solid line shows the signal from the run with Stycast constrain,
 dashed line - the signal after the pressure release. The shift
from zero degree for B phase can be due to Spin Waves
 modes in an inhomogeneous order parameter.}
 \label{pinning}
\end{figure}

We keep in mind the next scenario of aerogel density inhomogeneity
and anisotropy: There are a short scale averaging (about of a few
coherence lengthes),  middle scale (about a few micron, the healing
scale in $^3$He-A), long scale (about a few hundreds micron, the
healing length in $^3$He-B) and a global anisotropy. The first one
determines the $T_c$ shift and Imre-Ma state, described above. The
middle scale anisotropy averaging determines the local orientation
of order parameter in A-phase, and the long scale anisotropy
determines the local orientation of order parameter in B phase. The
global anisotropy determines the global orientation of the order
parameter. We can suppose that in aerogel the density anisotropy
oriented randomly. (If it was not oriented at the process of drying,
see\cite{Halp}) Its scattering decrease with increasing of
dimensions of averaging. After the first pressure treatment, which
reaches the plastic deformation limit, the global density
orientation became along the axis of pressure, but with a
significant random deviation in a perpendicular plain. The
additional pressure due to Stycast constrain, make orientation more
parallel to axis, but after its release the scattering restored.

We can characterize the shape of $^3$He-A NMR line in a local
approximation; that means that we attribute the local NMR frequency
to a local orientation of $\bf L$ which aline with the density
orientation averaged on a distance of texture healing length. In
Fig.4 one can see the deflection of $\bf L$ from vertical to
25$^\circ$ for this approximation. In the second run, which we have
made after Stycast pressure release, the deviation of $\bf L$ from
vertical orientation in A phase significantly increase. Take in mind
that Stylist gives an additional 1\% axial deformation, we can
estimate the global orientation without Stycast pressure on about
3\% and the local density anisotropy on the order of 1\% and
randomly oriented.

We can support this our vision by the data for B phase. The healing
length of the texture in B phase is much bigger then in A phase. The
averaging on large scale makes density random orientation very
small. That is why the B phase shows near vertical orientation of
averaged $\bf L$ for a both cases. The small deviation from zero
frequency shift can be explain by the well known shift due to Spin
Waves, which we can not take into account in our local
approximation.

We have performed many different rotation experiments. We have
cooled through  the transition under rotation or just rotated at
constant temperature or rotated with creation of the HPD. From these
experiments we conclude that influence of counterflow is very small
for the case of density anisotropy of the aerogel. The some
influence can be seen  if we will cross $T_c^a$ under rotation. In
this case the spin waves modes clearly seen in B phase
\cite{QFS2006}. The effect of counterflow, observed in \cite{Kyoto}
can also be seen but a very small amplitude. In A phase the
formation of magnetic coherent quantum state have been observed,
which we will discuss elsewhere. In brief, the aerogel in our case
is saturated by different types of topological defects, but
counterflow at $T_c^a$ can prevent the formation of some types of
topological defects, which makes order parameter more sensitive to
counterflow and to nonlinear NMR.

\section{Discussion and conclusions}
For to make a long story short,  we have demonstrated in a first
time, that the global anisotropy of aerogel  has a strong effect on
the orientation of orbital momentum in A and B phases. In both
phases the orbital momentum is oriented along the direction of
compression. The line broadening in $^3$He-A is bigger then in
$^3$He-B due to the difference of texture healing length.

We can put a new light on the problem of aerogel surface
orientation. It was found \cite{DmitOB} that the free surface of
aerogel orient $\bf L$ in B phase parallel to the surface. From our
results this effect can be explain by a local aerogel deformation at
the process of mechanical cutting of the sample. The mechanical
treatment can form a dense layer of aerogel near the surface, which
orient $\bf L$, not a boundary itself.

And finally, we have found that the Leggett frequency dependence
does not correspond to one suggested for A phase at the region of
temperature down to 0.9 Tc. The discrepancy is much larger than the
possible experimental error. Perhaps near Tc the A phase has some
kind of planar phase distortion as it takes place with A phase near
the wall!

We are thankful to N. Mulders for the aerogel sample and J. Pollanen
and G. Volovik for many discussions. The work was made in ISSP. The
collaboration with physicists from other institutions was supported
by Joint CNRS-JSPS project, by 21 century COE program and by KAKENHI
program (grant 17071009). Yu. Bunkov is particularly thankful for
ISSP hospitality.

\end{document}